\documentclass[11pt]{article}
\baselineskip
\usepackage{amssymb}
\usepackage{graphicx}
\usepackage{latexsym,cite}
\usepackage{epsfig}
\newcommand{\eq}[1]{\begin{equation}\label{#1}}
\newcommand{\en}{\end{equation}}
\newcommand{\ear}[1]{\begin{eqnarray}\label{#1}}
\newcommand{\rae}{\end{eqnarray}}
\newcommand{\ena}{\end{eqnarray}}
\newcommand{\beq}[1]{\begin{equation}\label{#1}}
\newcommand{\eeq}{\end{equation}}
\newcommand{\bea}[1]{\begin{eqnarray}\label{#1}}
\newcommand{\eea}{\end{eqnarray}}

\newcommand{\R}{\mathbb{R}}
\newcommand{\C}{\mathbb{C}}

\newcommand{\tr}{\mbox{tr}} 
\newcommand{\ide}{\mbox{1 \kern-.59em {\rm l}}}

\begin{document}
\begin{titlepage}
\vskip0.5cm
\begin{flushright}
DIAS-STP-06-11\\
\end{flushright}
\vskip0.5cm
\begin{center}
{\Large\bf Numerical simulations of a non-commutative theory: the scalar model on the fuzzy sphere}
\end{center}
\vskip1.3cm
\centerline{Marco~Panero}
\vskip1.5cm
\centerline{\sl  Institute for Theoretical Physics}
\centerline{\sl  University of Regensburg}
\centerline{\sl  93040 -- Regensburg}
\centerline{\sl  Germany}
\vskip0.5cm
\centerline{\sl  School of Theoretical Physics}
\centerline{\sl  Dublin Institute for Advanced Studies}
\centerline{\sl  10 Burlington Road}
\centerline{\sl  Dublin 4, Ireland}
\vskip0.5cm
\begin{center}
{\sl  e-mail:} \hskip 6mm \texttt{marco.panero@physik.uni-regensburg.de}
\end{center}
\vskip1.0cm

\begin{abstract}
We address a detailed non-perturbative numerical study of the scalar theory on the fuzzy sphere. We use a novel algorithm which strongly reduces the correlation problems in the matrix update process, and allows the investigation of different regimes of the model in a precise and reliable way. We study the modes associated to different momenta and the r\^ole they play in the ``striped phase'', pointing out a consistent interpretation which is corroborated by our data, and which sheds further light on the results obtained in some previous works. Next, we test a quantitative, non-trivial theoretical prediction for this model, which has been formulated in the literature: The existence of an eigenvalue sector characterised by a precise probability density, and the emergence of the phase transition associated with the opening of a gap around the origin in the eigenvalue distribution. 
The theoretical predictions are confirmed by our numerical results. Finally, we propose a possible method 
to detect numerically the non-commutative anomaly predicted in a one-loop perturbative analysis of the model, which is expected to induce a distortion of the dispersion relation on the fuzzy sphere. 
\end{abstract}
\end{titlepage}

\section{Introduction}\label{introsect}

The study of quantum field theory in non-commutative spaces has attracted considerable attention over the last years~\cite{Doplicher:1994tu, Landi:1997sh, Douglas:2001ba, Szabo:2006wx}. This research area has a long history, since the possibility of a quantised structure of spacetime at short distances was first mentioned as early as in the 1930's in some correspondence among Heisenberg, Peierls, Pauli and Oppenheimer~\cite{Pauli1,Pauli2}, and in the papers published by Snyder~\cite{Snyder:1946qz}, by Yang~\cite{Yang:1947ud} and by Moyal~\cite{Moyal:1949sk} during the 1940's. Although the original motivation to use non-commutativity as a tool to regularise QFT was soon frustrated --- while the renormalisation approach proved to be a  successful method to handle the divergences encountered in the formulation in commutative spacetime ---, non-commutative spaces have attracted renewed interest in more recent years, with the application of this formalism to solid-state physics, to the problem of the quantum Hall effect~\cite{Karabali:2004xq}, to the study of MHD waves in astrophysics~\cite{Bourouaine:2006mp}, and with the discovery of the relevance of such spaces to string theory~\cite{Connes:1997cr, Douglas:1997fm, Alekseev:1999bs, Seiberg:1999vs, Kar:1999tu, Kar:2000pb, Lizzi:2006te} and to a possible quantum theory of gravity~\cite{Chamseddine:1992yx, Lizzi:2002ib}. 

Groenewold-Moyal $\R^n_\theta$ spaces are among the most extensively studied non-commutative spaces. The properties of QFT defined in these spaces --- including those related to renormalisability, causality, non-locality, Poincar\'e invariance \emph{et c\oe tera} --- have been addressed in several works~\cite{Doplicher:1994zv, Doplicher:2001qt, Filk:1996dm, Chen:2001qg, Jain:2003xs, Chaichian:2004za, Szabo:2001kg, Minwalla:1999px, Hinchliffe:2002km, Gubser:2000cd, Castorina:2003zv, Langmann:2002cc, Steinacker:2005wj, Grosse:2004yu, Rivasseau:2005bh, Govindarajan:2006vh, Grosse:2005ig, Lizzi:2006xi, Stern:2006zt, Grosse:2006qv, Gracia-Bondia:2006yj, Grosse:2006tc, Disertori:2006uy, Grosse:2007dm}; a lattice-like regularised formulation has also allowed to investigate numerically various aspects of these models~\cite{Ambjorn:1999ts, Ambjorn:2000nb, Ambjorn:2000cs, Azuma:2004zq, Bietenholz:2002ch, Bietenholz:2002vj, Ambjorn:2002nj, Bietenholz:2004xs, Bietenholz:2005iz, Bietenholz:2006cz}. In particular, one of the most interesting --- albeit troublesome --- features is the fact that the effective action describing QFT in a Groenewold-Moyal space is divergent when the external momentum along the non-commutative directions vanishes: this effect arises from the integration of the high-energy modes in non-planar loop diagrams, and is henceforth called ``ultra-violet/infra-red (UV/IR) mixing''.

Another class of non-commutative spaces is given by fuzzy spaces: they are built approximating the infinite-dimensional algebra of functions on some particular manifold by means of a finite-dimensional algebra of matrices. Under some conditions, this construction is possible for even-dimensional co-adjoint orbits of Lie groups which are symplectic manifolds --- see~\cite{Madore:1991bw, Alexanian:2001qj, Balachandran:2001dd, Hammou:2001cc, Medina:2002pc, Grosse:1995pr, GrosseReiter, Balachandran:2002jf, Vaidya:2003ew, Dolan:2003kq, Lizzi:2003ru, Lizzi:2005zx, Balachandran:2005ew, Sheikh-Jabbari:2006bj} and references therein. In particular, co-adjoint orbits of semi-simple Lie groups are adjoint orbits; examples include the $\C P^n$ complex projective spaces. The most-widely known example of a fuzzy space is the fuzzy two-sphere $S^2_F$~\cite{Madore:1991bw}, built truncating the algebra of functions on the commutative sphere $S^2$ to a maximum angular momentum $l_{\mbox{\tiny{max}}}$. The fuzzy sphere depends on two parameters: the matrix size $N=l_{\mbox{\tiny{max}}} +1$ and the radius $R$; the commutative sphere and the non-commutative plane can be obtained in different limits of $N$ and $R$. 

A one-loop perturbative calculation shows that, for every finite $N$, QFT on the fuzzy sphere is finite; furthermore, the theory is not affected by the UV/IR mixing problem~\cite{Chu:2001xi} (although --- due to a non-commutative anomaly --- the latter re-emerges once a double-scaling limit is taken, in which the fuzzy sphere goes over to the non-commutative plane $\R^2_\theta$). This feature, as well as the fact that the fuzzy approach explicitly preserves the symmetries of the original manifold for any value of $N$ and allows a well-defined treatment of the topological properties~\cite{Baez:1998he, Grosse:1995jt, Balachandran:1999hx, Presnajder:1999ky, Carow-Watamura:1998jn, Steinacker:2003sd, Behr:2005wp, Nair:2006qg, Valtancoli:2006dw, Dolan:2006tx, Murray:2006pi, Saemann:2006gf, Aoki:2006wv,  Delgadillo-Blando:2006dp,  Steinacker:2007iq}, has led to suggest the fuzzy space as a potentially interesting candidate for regularisation of quantum field theory. 

As a matter of fact, QFT on the fuzzy sphere is mathematically well-defined and finite~\cite{Grosse:1995ar}, and the formulation is amenable to a non-perturbative approach and to numerical studies using Monte Carlo simulations, with fields represented as finite-dimensional matrices. This approach has been followed in various recent works~\cite{Martin:2004un, GarciaFlores:2005xc, Medina:2005su, Azuma:2004yg, Anagnostopoulos:2005cy, Azuma:2005pm, O'Connor:2006wv}.

In the present paper we address a detailed Monte Carlo study of the $\Phi^4$ scalar field theory on the fuzzy sphere; among other issues of interest, this simple model provides a laboratory to test the possibility to use the fuzzy space approach as a potential regularisation scheme for more realistic field theories. As it concerns the practical implementation of Monte Carlo simulations of the model, we shall present a novel algorithm, which reduces the autocorrelation time, combining overrelaxation steps with ergodic configuration updates.

Preliminary results of this study have been presented in~\cite{Panero:2006cs}.

This manuscript has the following structure: in section~\ref{theorysect} the theoretical framework underlying the model is recalled, and the basic notations are introduced. In section~\ref{resultsect} we discuss the implementation of the numerical simulations of the model, and present the results obtained with our algorithm. In section~\ref{discussionsect} we comment on the implications of these results
and on possible research perspectives. A technical discussion of the algorithm is presented in the Appendix~\ref{overrelsect}.

\section{Review of the construction of the model}\label{theorysect}

A general discussion of the mathematical construction of fuzzy spaces can be found in many excellent articles and books, like, for instance~\cite{Balachandran:2005ew}; for the scalar field theory on the fuzzy sphere $S^2_F$, we refer the reader to the detailed presentation in~\cite{Grosse:1995ar}.

The basic idea is to replace the infinite-dimensional, commutative algebra of polynomials generated by the $\{ x_i \}_{i=1,2,3}$ coordinates on the two-dimensional sphere $x_i x_i = R^2$ embedded in $\R^3$ with a non-commutative algebra generated by $\{ \hat{x}_i \}_{i=1,2,3}$ operators satisfying a (trivially rescaled) $su(2)$ Lie algebra. 
The latter
can be realised using the Wigner-Jordan construction of the $su(2)$ generators, restricting to the finite-dimensional ($N$-dimensional) subspace of the Fock space generated by a pair of mutually commuting creation operators. Accordingly, the algebra of functions on the commutative sphere $S^2$ is replaced by the Mat$_N$ algebra, whose elements can be expanded into irreducible representations of $su(2)$.

The fuzzy sphere admits the commutative sphere $S^2$ and the Groenewold-Moyal plane $\R^2_\theta$ as two different limits: the former is recovered for $N \rightarrow \infty$ with $R$ fixed, whereas the latter can be obtained --- at least locally --- via a stereographic projection from a fixed point, in the double limit: $N \rightarrow \infty$, $R \rightarrow \infty$, keeping $R^2/N$ fixed. 

The action for a massive, neutral, scalar field with quartic interactions on the fuzzy can be defined --- according to the conventions used in~\cite{Martin:2004un} --- as:
\eq{action}
S= \frac{4\pi }{N} \tr \left( \Phi \left[ L_i , \left[ L_i, \Phi \right] \right] + r R^2 \Phi^2 + \lambda R^2 \Phi^4 \right) \; ,
\en
where $\Phi \in$ Mat$_N$ is hermitian and 
can be expanded in the $\{ \hat{Y}_{l,m} \}$ polarisation tensor basis as:
\eq{phiexpansion}
\Phi=\sum_{l=0}^{N-1} \sum_{m=-l}^{l} c_{l,m} \hat{Y}_{l,m} \; .
\en

The model can be quantised in the path integral approach, defining the expectation values of generic observables $\mathcal{O}=\mathcal{O}(c_{l,m})$, which can be evaluated perturbatively, or estimated numerically from Monte Carlo simulations. 

The perturbative treatment of the theory is formulated via a proper definition of the Feynman rules --- in particular, the interaction vertices are modified in a non-trivial way, which depends on $N$, and is consistent with the fusion rules. A careful one-loop analysis of this model~\cite{Chu:2001xi} shows that the UV/IR mixing phenomenon does not occur on $S^2_F$; however, a non-commutative anomaly shows up, as a finite difference between planar and non-planar tadpole diagrams. 

This anomaly is expressed by a rotationally invariant, non-local contribution to the quantum effective action; it  distorts the dispersion relation, and 
survives the limit to the commutative sphere. This seriously threatens the possibility to consider the fuzzy approach as a \emph{bona fide} regularisation scheme for theories defined in a commutative space; however, according to~\cite{Dolan:2001gn}, the problem may be overcome, redefining the interaction term in the matrix action with a normal-ordering prescription, which allows to cancel the undesired momentum-dependent quadratic terms in the effective action. An alternative, and more general, possibility would be to include rotationally symmetric higher derivative terms in the action, as suggested in~\cite{briandenjoe}.

On the other hand, when the Groenewold-Moyal plane limit is taken, the non-commutative anomaly reproduces the logarithmic divergence which is characteristic of the UV/IR mixing~\cite{Minwalla:1999px}.

Fuzzy spaces were studied in~\cite{Steinacker:2005wj} as a mean to regularise scalar field theory in non-commutative $\R_\theta^n$ spaces. There, it was shown that different phases can be distinguished in the large $N$ limit, according to the form of the distribution of eigenvalues associated to the matrix $\Phi$. The argument goes as follows: For the free case, in the large $N$ limit and assuming that the cut-off is much larger than the non-commutative scale, the leading contribution to the expectation values of 
even powers ($2k$) of the field come from planar diagrams. 
Next, one observes that, in the large $N$ limit, a clustering property  for the expectation values of products of integrals of the field holds, and  
implies that the measure is strongly localised. 
This allows to identify the eigenvalue distribution as the Wigner semi-circle law, which
holds for gaussian random matrix models~\cite{Mehta,Stephanov:2005ff}. 

Next, one can generalise to include interactions: for a 
scalar theory with bare square mass $m^2$ and quartic interactions with coupling $\frac{g}{4}$ in non-commutative space
the resulting eigenvalue distribution is known exactly~\cite{Brezin:1977sv}: 
for $m^2$ larger than a critical value $m^2_{\mbox{\tiny{crit}}}$ 
the eigenvalue density $\rho_g (\varphi)$ has a connected support and is given by:
\eq{onecutdistribution}
\rho_g (\varphi)= \frac{1}{2\pi} \left[ g^\prime \left( \varphi^2 +\frac{1}{2} \right) + (m^\prime )^2 \right] \sqrt{1-\varphi^2}
\en
where $g^\prime$ and $(m^\prime )^2$ are related to $g$ and $m^2$ --- see~\cite{Steinacker:2005wj} for the details.
On the contrary, for $m^2 < m^2_{\mbox{\tiny{crit}}}$, the eigenvalue distribution exhibits two disconnected peaks of finite width.

Although the arguments underlying this derivation are expected to hold only approximately in the two-dimensional case (because the dominance of the planar diagrams over the non-planar ones is weaker than in four dimensions), one can check if the numerical results are consistent with the predicted features.

The real scalar model with quartic interactions on the fuzzy sphere was studied numerically in~\cite{Martin:2004un, GarciaFlores:2005xc}, where it was shown that the model exhibits three different phases: a disordered phase, in which the field typically fluctuates around zero; a uniform order phase, characterised by fluctuations around the broken-symmetry minima of the potential, and a non-uniform order phase, which was described as new, intermediate, phase, intrinsically related to the matrix nature of the fuzzy regularisation. The appearance of the latter was interpreted assuming that, in a certain parameter range, the kinetic contribution to the action might be negligible, and the dynamics of the system were effectively reduced to the framework of a pure potential model~\cite{Shimamune:1981qf, Bleher:2002ys}. 
This phase was also described as analogous to the striped phase predicted in~\cite{Gubser:2000cd} for non-commutative Groenewold-Moyal spaces, and observed numerically in~\cite{Ambjorn:2002nj, Bietenholz:2004xs}.

\section{Numerical simulations}
\label{resultsect}

The numerical approach to the model is completely straightforward, and, under many respects, analogous to the more conventional lattice setting.\footnote{It should be noted, however, that the fuzzy space approach to a quantum theory differs with respect to the lattice formulation in a number of aspects, which also have practical implications for the computer simulations. In particular, in lattice field theory there exist many efficient update methods (including, for instance, those described in~\cite{Kennedy:1985nu, Adler:1981sn, Brown:1987rr, Fabricius:1984wp, Luscher:2001up, Zach:1997yz, Panero:2005iu}) which are based on the locality of the discretised action; these methods allow to strongly damp the autocorrelation among subsequent configurations in the Markov chain. Furthermore, parallel computation can often be implemented in a straightforward way. On the contrary, in the fuzzy setting, the dynamics of each degree of freedom is non-trivially entangled with each other's, and --- in general --- the implementation of analogous techniques is not trivial.} Expectation values of the observables are estimated from averages over finite ensembles of matrices $\{ \Phi \}$, characterised by a statistical weight which depends on the model dynamics; the algorithm generating the matrix ensemble is built combining overrelaxation steps~\cite{Adler:1981sn, Brown:1987rr} and canonical updates.

For the configuration-updating process, different types of pseudo-random number generators were compared; the G05CAF generator of the NAG library has eventually been used in the production runs.

For each choice of the $(N, r, \lambda, R)$ parameters, the autocorrelation time between elements in the thermalised matrix ensemble has been calculated using the auto-windowing procedure~\cite{Madras:1988ei}; the expectation values of the various observables have been evaluated from ensembles of statistically uncorrelated matrices. The data analysis was done using standard techniques, and errorbars have been estimated using the jackknife method --- see, for instance,~\cite{shaotubook}.

In the $r<0$ regime, it is particularly interesting to study the behaviour associated with the various $l$-modes. The 
classical minima of the potential 
correspond to uniform distributions and obviously their physical content is purely described by the scalar ($l$=0) channel. Since we are dealing with a quantum model and the system size is finite, the ground state is actually unique, as quantum fluctuations allow finite-action tunneling events between the two minima. When $r$ is negative in sign and large in modulus, the profile of the potential is very steep, and the typical matrix configurations lie in a close neighbourhood around the classical minima; the trace of $\Phi$ allows to identify around which of the two minima the matrix is lying at a given Monte Carlo time. The ``trace susceptibility'', defined as: 
\eq{tracesusceptibilitydef}
\chi = \langle \left( \tr \Phi \right)^2 \rangle - \langle \left| \tr \Phi \right| \rangle^2 \; ,
\en
encodes the physical information about the fluctuations of $\tr \Phi$. When $r$ is increased to values closer to zero, the trace susceptibility exhibits a peak, corresponding to a maximum in the quantum fluctuations. As usual, the location of a maximum in the susceptibility approximately\footnote{Modulo corrections due to the finiteness of $N$ and/or $R$.} identifies the critical value where the ``phase transition'' to the disordered phase would occur, for an infinite system.

\begin{figure}[htbp]
\centering
\epsfig{file=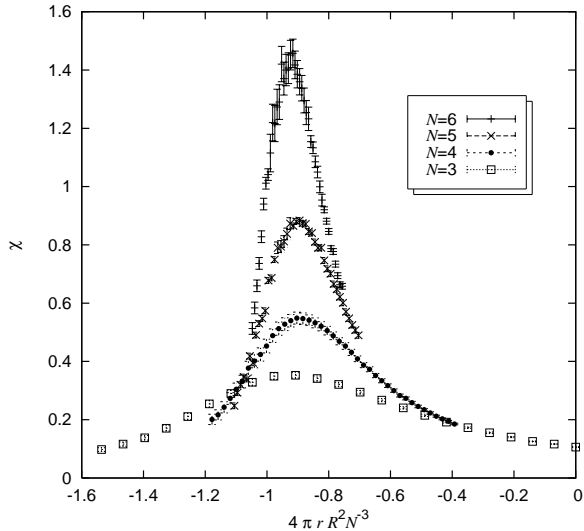, width=0.5\textwidth}
\caption{Behaviour of the trace susceptibility, as defined in eq.~(\ref{tracesusceptibilitydef}).}
\label{susceptibility_fig}
\end{figure}
The model already exhibits good scaling properties for small matrix sizes $N < 10$. As an example, figure~\ref{susceptibility_fig} shows the results obtained for the susceptibility, along the $\lambda=\frac{N^3}{4 \pi}$ line: the peak gets very pronounced as $N$ is increased, and its location is not affected by strong finite-$N$ effects.

Taking a closer look at the matrix ensembles and at their physical content in  
non-zero momenta 
allows to detect the 
non-uniform 
modes, and the r\^ole they play in the regime under consideration. A simple variational analysis shows that also the commutative sphere can admit non-uniform configurations characterised by a finite, negative total amount of action.\footnote{It is easy to verify analytically that --- at least in some parameter ranges --- even an axially symmetric (but non-uniform) configuration described in terms of the first Legendre polynomial may be favoured with respect to the $\Phi=0$ uniform configuration, and thus mediate the tunneling events among the classical minima of the potential.} Numerically, the relevance of these configurations can be quantified through the average values of the square moduli of the $c_{l,m}$ coefficients for $l \ge 1$. The latter are indeed found to be non-vanishing for all of the $N$, $\lambda$ and $r$ values which we investigated; again, in agreement with~\cite{Martin:2004un, GarciaFlores:2005xc}, we observed a parameter range where the expectation value of the $l=1$ mode is larger than the scalar component.

The transition from the uniform- to the non-uniform-order phase can be 
interpreted as an effect arising when the kinetic contribution to the action becomes negligible with respect to the potential one~\cite{Martin:2004un, GarciaFlores:2005xc}. 

\begin{figure}[htbp]
\centering
\epsfig{file=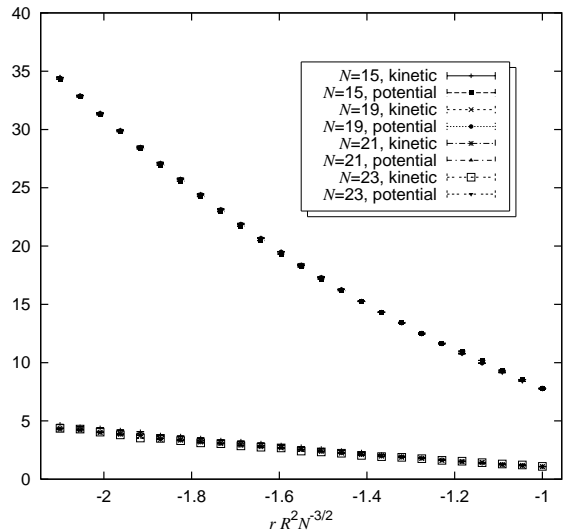, width=0.5\textwidth}
\caption{Average values of the kinetic term and modulus of the potential per degree of freedom, in the proximity of the transition between the disorder and striped phases. Data obtained at fixed $\lambda = \frac{3 \pi N}{25 R^2}$.}
\label{potkin_fig}
\end{figure}

As figure~\ref{potkin_fig} shows, our results confirm that at the transition the average kinetic contribution is much smaller than the modulus of the potential. However, this behaviour persists (and is, in fact, enhanced) down to more strongly negative $r$-values in the uniform order regime, too --- a fact that does not allow to interpret the existence of the striped phase as solely due to the potential dominance.

Rather, the results are compatible with the fact that 
the tunneling between the minima of the potential may be mediated by matrix configurations corresponding to non-spherically-symmetric distributions on the sphere. This can be easily justified from the analytical point of view (evaluating explicitly the action associated to such configurations), and is fully consistent with the numerical data (which confirm a non-negligible expectation value for the modes with $l \neq 0$ when $r$ is not very far from zero). 

Having presented the general features of the phase structure of the model, we now address the 
test of the theoretical predictions formulated in~\cite{Steinacker:2005wj}\footnote{Note that the conventions in~\cite{Steinacker:2005wj} differ with respect to eq.~(\ref{action}) and to those used in~\cite{Martin:2004un} by a trivial rescaling of the $\Phi$ matrix, and the coefficients of the quadratic and quartic terms in the potential are denoted as $m^2$ and $g$, respectively.}: The model can be described by means of random matrix methods, and is characterised by well-defined properties of the eigenvalue sector. In particular, the phase transition is associated with a change in the topology of the support of the eigenvalue distribution.


We concentrated our attention onto matrices of size $N = 15$, $19$, $21$, $23$, $31$, $41$ and $81$.

\begin{figure}
\centerline{\includegraphics[width=0.5\textwidth]{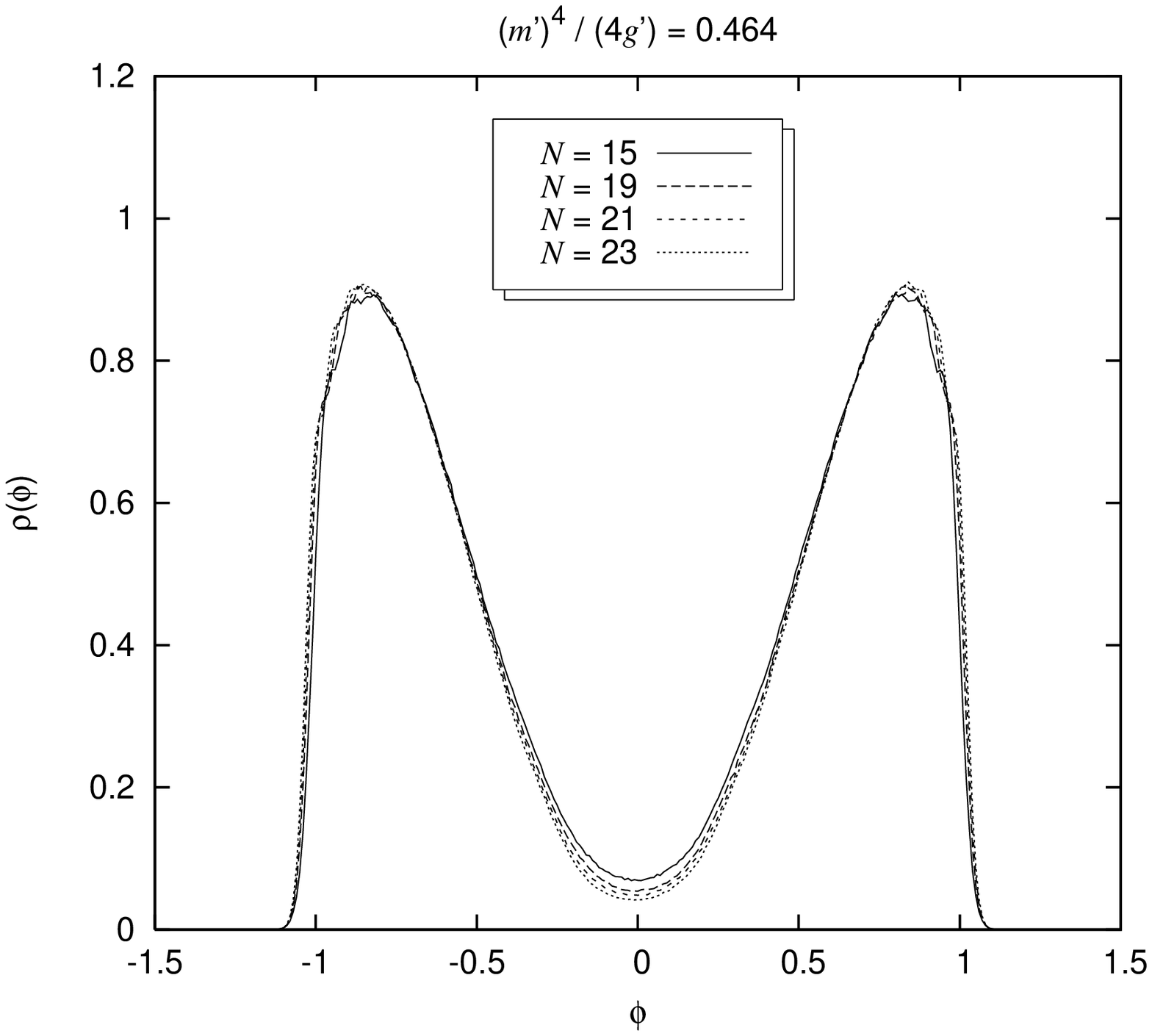}\includegraphics[width=0.5\textwidth]{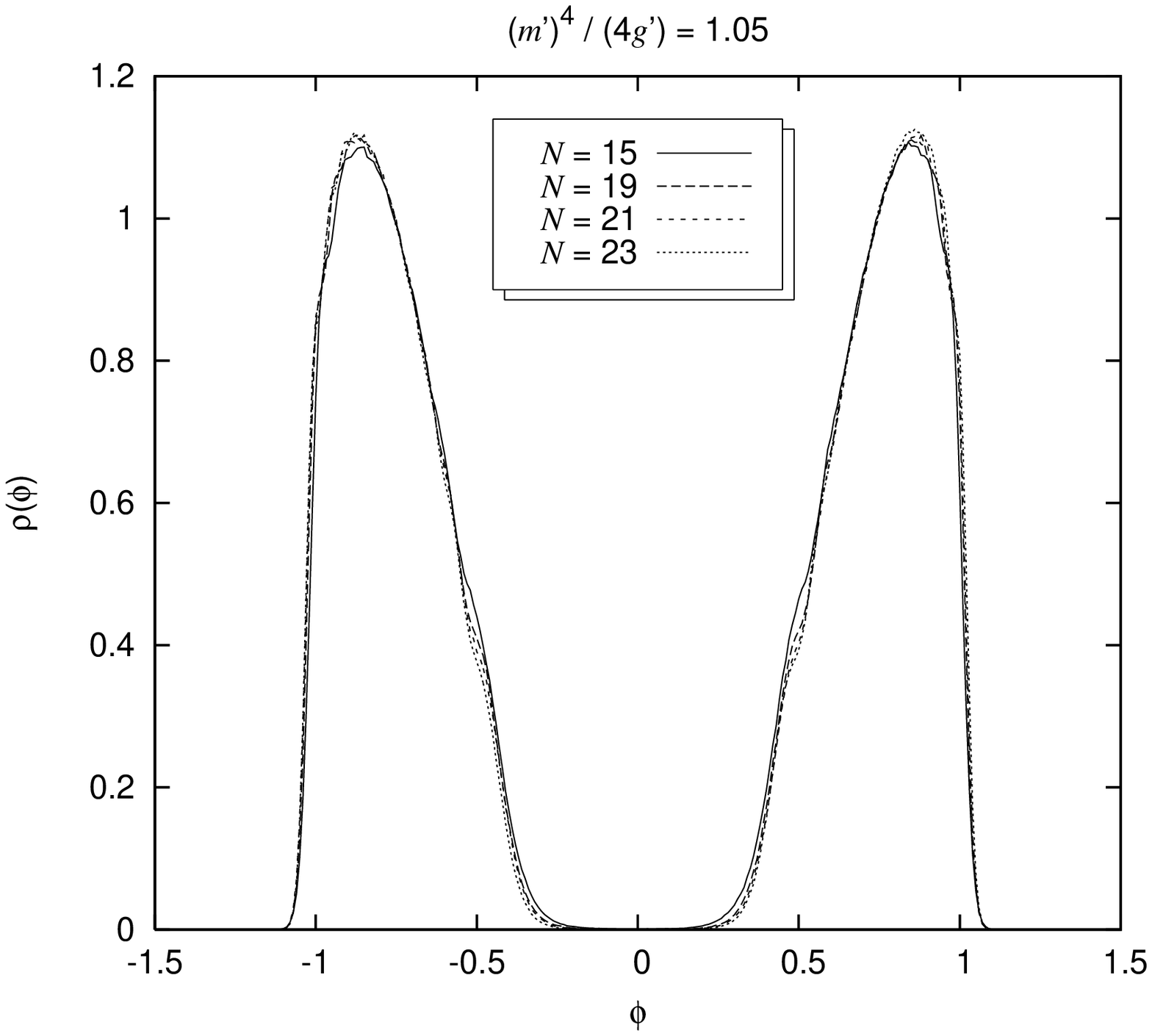}}
\centerline{\includegraphics[width=0.5\textwidth]{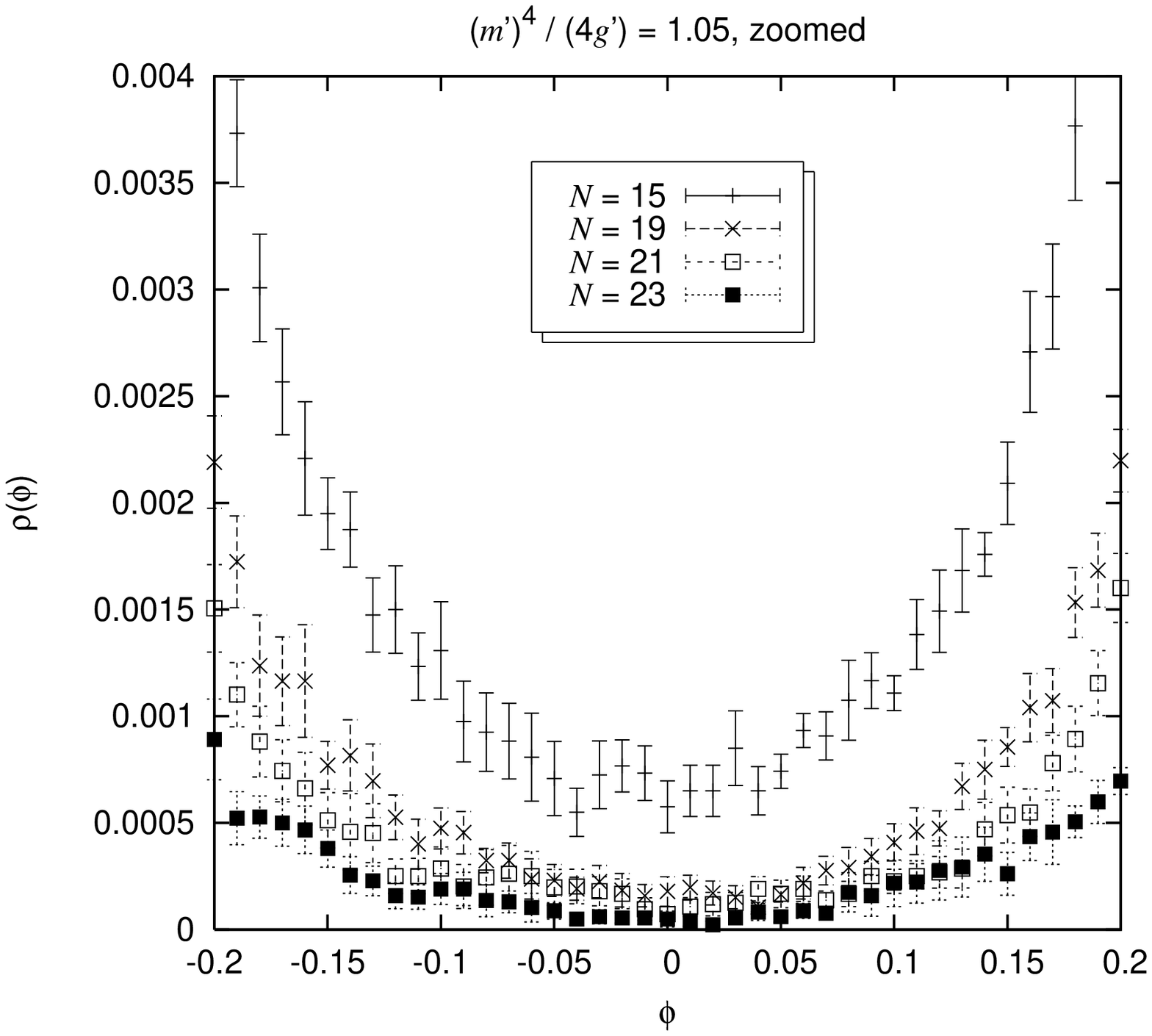}\includegraphics[width=0.5\textwidth]{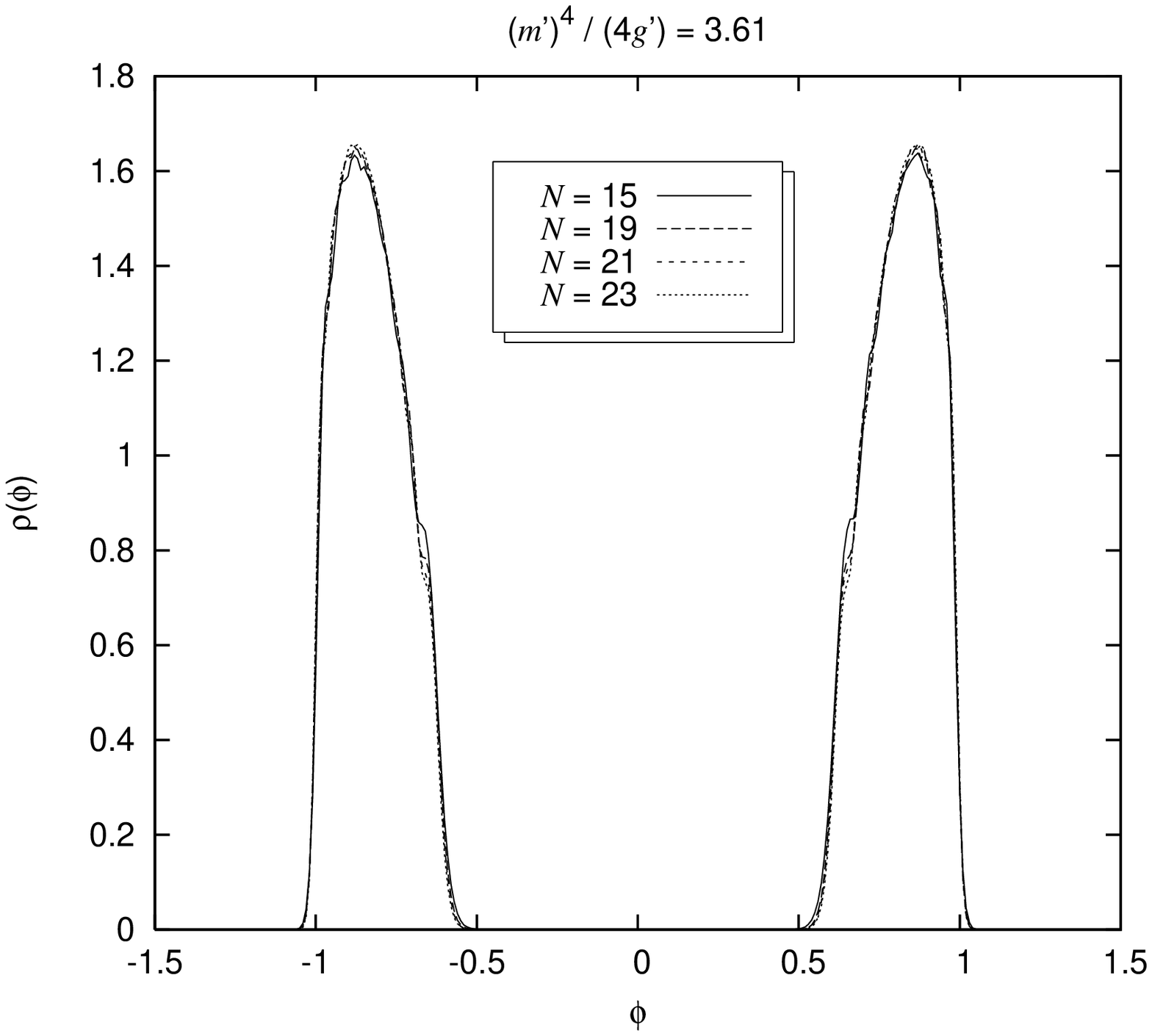}}
\caption{Evolution of the eigenvalue distribution from the one-cut (top-left plot) to the two-cut phase (bottom-right). The theoretical prediction for the transition point is: $(m')^4 / (4g') = 1$.}
\label{transition_fig}
\end{figure}

At intermediate and large values of $g$ the data agree well with the theoretical prediction, and the theoretical prediction is indeed confirmed in the double-scaling limit. Figure~\ref{transition_fig} shows the evolution of the eigenvalue density, from the one-cut to the two-cut phase.

We should emphasise that the agreement observed is highly non-trivial, because the arguments underlying the derivation in~\cite{Steinacker:2005wj} are expected to hold only approximately in two dimensions.\footnote{This is due to the weaker dominance of planar diagrams over the non-planar ones in $D=2$.} Although it would not be completely justified to perform a fit of the data to the curve, 
the agreement between the numerical results and the theoretical curve, which is a completely parameter-free prediction and \emph{does not} involve any fitting procedure, is striking.

On the contrary, the model exhibits significant quantitative deviations from the expected critical point at small $g$ values. Although the qualitative pattern of the transition is confirmed, the location of the transition point is shifted towards more negative $r$ values. In fact, this is not surprising: in this regime, the secondary maxima in the eigenvalue distribution are no longer negligible; the system could rather be described by a multi-trace model~\cite{christiananddenjoeinpreparation}.

However, the general agreement of numerical results and theoretical predictions is also confirmed by the scaling properties of the eigenvalue density support. The samples where the observed width of the distribution support differs from the expected one by an amount of the order of $5$ -- $10 \% $ or more correspond to strongly negative $r$ values --- which actually may even already lie in the ``uniformly ordered'' phase.\footnote{The critical line corresponding to the transition from the non-uniform to the uniform order phase was obtained numerically in~\cite{GarciaFlores:2005xc}, but a full theoretical description of this transition in the fuzzy sphere setting is still missing.}

Table~\ref{critical_r_comparison_tab} shows results for the numerical critical values of $r R^2 $, for various matrix sizes.

    \begin{table}[h]
    \begin{center}
    \begin{tabular}{|c|c|c|c||c|c|c|c|}
    \hline
    $g R^2$ & $N$ & $(r R^2)_{\mbox{\tiny{crit}}}^{\mbox{\tiny{num}}}$ & $(r R^2 )_{\mbox{\tiny{crit}}}^{\mbox{\tiny{th}}}$ & $g R^2$ & $N$ & $(r R^2)_{\mbox{\tiny{crit}}}^{\mbox{\tiny{num}}}$ & $(r R^2 )_{\mbox{\tiny{crit}}}^{\mbox{\tiny{th}}}$ \\
    \hline
    \hline
         $  8 \pi N $ & $ 15 $ & $  -258.1 \pm  5.2 $ & $    -230.6 $ &   $\frac{6}{5} \pi N $ & $ 15 $ & $  -141.6 \pm  4.4 $ &  $  -94.9 $ \\ 
                      & $ 19 $ & $  -353.0 \pm  7.4 $ & $    -333.7 $ &                        & $ 19 $ & $  -198.8 \pm  7.3 $ &  $ -135.6 $ \\ 
                      & $ 21 $ & $  -410.2 \pm  8.6 $ & $    -390.0 $ &                        & $ 21 $ & $  -241.8 \pm  8.8 $ &  $ -157.7 $ \\ 
                      & $ 23 $ & $  -460.3 \pm  9.9 $ & $    -449.2 $ &                        & $ 23 $ & $  -268.9 \pm  9.6 $ &  $ -180.9 $ \\ 
    \hline                                                                                                                                   
         $  4 \pi N $ & $ 15 $ & $  -194.9 \pm  7.6 $ & $    -169.2 $ &   $ \frac{4}{5} \pi N $ & $ 15 $ & $ -126.4 \pm  5.5 $ &   $  -77.6 $  \\ 
                      & $ 19 $ & $  -277.9 \pm  9.8 $ & $    -243.4 $ &                         & $ 19 $ & $ -183.2 \pm  6.2 $ &   $ -110.9 $  \\ 
                      & $ 21 $ & $  -323   \pm 13   $ & $    -283.8 $ &                         & $ 21 $ & $ -220.1 \pm  7.2 $ &   $ -128.9 $  \\ 
                      & $ 23 $ & $  -370   \pm 14   $ & $    -326.2 $ &                         & $ 23 $ & $ -252.3 \pm  8.3 $ &   $ -147.8 $  \\ 
    \hline
         $  2 \pi N $ & $ 15 $ & $  -134.5 \pm  6.4 $ & $    -121.9 $ &   $ \frac{3}{5} \pi N $ & $ 15 $ & $ -117.9 \pm  3.4 $ &   $  -67.3 $  \\  
                      & $ 19 $ & $  -194.8 \pm  8.9 $ & $    -174.6 $ &                         & $ 19 $ & $ -172.9 \pm  4.8 $ &   $  -96.0 $  \\  
                      & $ 21 $ & $  -219.3 \pm  9.1 $ & $    -203.1 $ &                         & $ 21 $ & $ -206.5 \pm  5.6 $ &   $ -111.6 $  \\  
                      & $ 23 $ & $  -280   \pm  13 $  & $    -233.0 $ &                         & $ 23 $ & $ -243.1 \pm  6.4 $ &   $ -128.0 $  \\  

    \hline
          $  \pi N $ & $ 15 $ & $ -132.9   \pm 7.4 $  & $  -86.7 $    &   $ \frac{\pi}{5}  N $  & $ 15 $ & $  -99.7 \pm 3.7 $ &   $  -38.9 $  \\ 
                     & $ 19 $ & $ -189.5   \pm 9.2 $  & $ -123.9 $    &                         & $ 19 $ & $ -143.4 \pm 4.1 $ &   $  -55.5 $  \\ 
                     & $ 21 $ & $ -235     \pm 11  $  & $ -144.0 $    &                         & $ 21 $ & $ -168.0 \pm 6.3 $ &   $  -64.5 $  \\ 
                     & $ 23 $ & $ -261     \pm 15  $  & $ -165.2 $    &                         & $ 23 $ & $ -213.5 \pm 8.7 $ &   $  -73.9 $  \\ 
    \hline                    
         $ 10 \pi N $ & $ 15 $ & $  -285.1 \pm  5.1 $ & $    -254.0 $ &    $\frac{8}{5} \pi N $ & $ 15 $ & $  -152.5 \pm  4.5 $ & $ -109.4 $  \\ 
                      & $ 19 $ & $  -392.0 \pm  7.2 $ & $    -368.0 $ &                         & $ 19 $ & $  -214.3 \pm  6.2 $ & $ -156.4 $  \\ 
                      & $ 21 $ & $  -442.9 \pm  8.4 $ & $    -430.4 $ &                         & $ 21 $ & $  -256.2 \pm  7.2 $ & $ -181.9 $  \\ 
                      & $ 23 $ & $  -507.7 \pm  9.6 $ & $    -496.1 $ &                         & $ 23 $ & $  -285.4 \pm  8.3 $ & $ -208.7 $  \\ 
                      & $ 31 $ & $  -803   \pm   25 $ & $    -789.1 $ &                         &  &  &  \\ 
                      & $ 41 $ & $  -1262  \pm   54 $ & $   -1215.1 $ &                         &  &  &  \\ 
                      & $ 81 $ & $  -3680  \pm  290 $ & $   -3435.4 $ &                         &  &  &  \\ 
    \hline
    \end{tabular}
    \end{center}
    \caption{Numerical and theoretical values for the quadratic coupling at the critical point.}
    \label{critical_r_comparison_tab}
    \end{table}

In the remaining part of this section, we 
shortly discuss 
the case when the classical potential has a unique minimum, and propose a method to detect the effect of the non-commutative anomaly. 

A perturbative study of the model in this regime was presented in~\cite{Chu:2001xi}\footnote{In the notations of~\cite{Chu:2001xi}, the radius is set to the unit, the $\Phi$ matrix (whose size is denoted as $N+1$) is rescaled by a factor $\sqrt{2}$ with respect to eq.~(\ref{action}), and the coefficients of the quadratic and quartic terms in the potential are denoted as $\mu^2$ and $\frac{g}{3!}$, respectively.}. As it was discussed above, the non-commutative anomaly shows up as a (mild) non-local effect, distorting the energy-momentum relation on the fuzzy sphere by a finite amount.

The 
one-loop effective action on the fuzzy sphere is~\cite{Chu:2001xi}:
\eq{effectiveaction}
S_{\mbox{\tiny{one-loop}}} = S_0 + \frac{4 \pi}{N+1} \tr \left[ \frac{\delta \mu^2}{2} \Phi^2 - \frac{g}{24 \pi}\Phi  h \left( \tilde \Delta \right) \Phi \right] + O\left( \frac{1}{N}\right) \; ,
\en
where $\delta \mu^2$ is the square-mass renormalisation:
\eq{squaremassrenormalisation}
\delta \mu^2= \frac{g}{8 \pi} \sum_{J=0}^{N} \frac{2J+1}{J(J+1) +\mu^2} \; ,
\en
while the non-commutative anomaly is given by the contribution involving $h\left( \tilde \Delta \right)$; $h(x)$ is the harmonic number: $h(x)=\sum_{t=1}^x \frac{1}{t}$, with $h(0)=0$, and $\tilde \Delta $ is a function of the Laplace operator, whose eigenvalue when acting on $\hat{Y}_{l,m}$ is $l$. In particular, the action of $h \left(\tilde \Delta \right) $ on the $\Phi$ matrix appearing in  eq.~(\ref{phiexpansion}) reads:
\eq{htildedeltaaction}
h \left( \tilde \Delta \right) \Phi = \sum_{l=1}^{N-1} \left[ \left( \sum_{t=1}^l \frac{1}{t} \right) \sum_{m=-l}^{l} c_{l,m} \hat{Y}_{l,m} \right] \; .
\en 
In order to 
disentangle among the various terms contributing to the effective action, one can tune the parameter in such a way, that the main momentum-dependent contribution at order $g$ comes from the non-commutative anomaly term only. Then, the 
spectrum of relative weights associated with the various spin channels gets distorted, as $g$ is changed: 
larger probabilities are expected for higher $l$ channels, when $g$ is increased.

\begin{figure}[htbp]
\centering{\epsfig{file=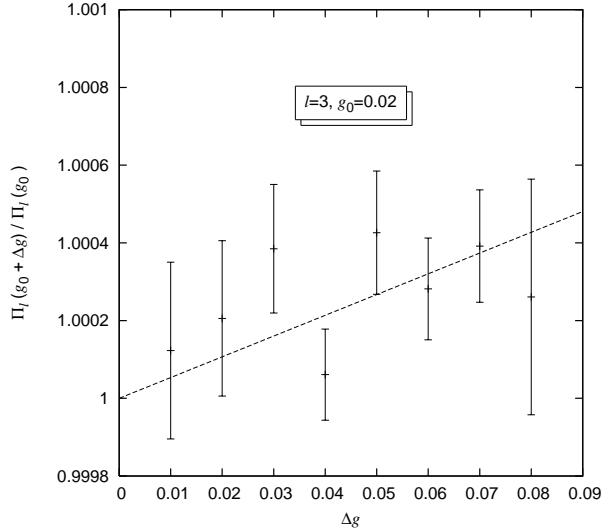, width=0.5\textwidth}}
\caption{At large $\mu$ values, the channels associated to non-vanishing momenta are expected to be slightly enhanced by the non-commutative anomaly. The plot shows preliminary results from tests over an ensemble of matrices of size $N+1=7$, at $\mu^2 = 200$, in comparison with the theoretical prediction (dashed line).}
\label{nca_fig}
\end{figure}

Preliminary numerical tests (see figure~\ref{nca_fig} as an example) confirm that the scalar component does not depend on $g$, 
while 
$\Pi_l(g)$, the power of the modes associated to higher $l$ values, which can be defined in terms of the coefficients appearing on the right-hand side of eq.~(\ref{phiexpansion}) as:
\eq{power}
\Pi_l = \langle \sum_{m=-l}^{l} | c_{l,m} |^2 \rangle \; ,
\en 
slowly increases with $g$, in rough agreement with the theoretical expectation.
However, due to the theoretical approximations and technical difficulties involved in the observation of these fine effects, our data do not allow to make conclusive statements about this issue. This problem may be addressed more thoroughly in future work.

\section{Discussion}\label{discussionsect}

The data in the previous section confirm the theoretical predictions for a scalar theory on the fuzzy sphere, as well as previous numerical results for the same model. The novel algorithm we used for the simulation proved to be very efficient, enabling us to obtain high-precision results for both the regimes that were investigated. 
The algorithm strongly reduces the correlation among subsequent configurations in a Markov chain, in a way which is compatible with the dynamics of the quantum system, and limiting undesired numerical artefacts to a minimum.

As it concerns the case of a classical scalar potential with two degenerate minima, our data, obtained from simulations at a large number of points in the space of physical parameters of the model, confirm and generalise the results obtained in similar numerical works~\cite{Martin:2004un, GarciaFlores:2005xc}. We have discussed the r\^ole of non-uniform configurations on a fixed, finite radius sphere and their relevance to 
tunneling events connecting the classical vacua --- 
a phenomenon which is not due to the non-commutative nature of the fuzzy regularisation, nor can it be directly interpreted as a signature of the UV/IR mixing.\footnote{It may be instructive to point out that the situation for a finite-radius sphere is different with respect to the case of an infinite plane.}

Yet, the relation to UV/IR mixing shows up, once one considers the double-scaling limit, and looks at the distribution for the matrix eigenvalues, which 
behave as a set of collective, intrinsically non-local degrees of freedom, and whose statistical properties can be worked out via random matrix methods. The UV/IR mixing manifests itself as the high-energy modes suppress the distribution of the low-energy ones. 

In our numerical study, we have successfully compared the observed eigenvalue distribution with the predicted pattern. 
The eigenvalues, rescaled through a factor predicted by the theory, fit very well to the $[-1,1]$ range, and their density undergoes a transition from the ``one-cut'' to the ``two-cut'' regime. Although for the two-dimensional case one would only expect an approximate agreement, the data for $\rho(\varphi)$ follow the theoretical curve very closely --- except at very small $g$-values, where a multi-trace model~\cite{christiananddenjoeinpreparation} would probably provide a better description of the system.

Next, we have also considered the $m^2 > 0$ regime, and proposed an approximate method to observe the non-commutative anomaly, through the spectrum distortion at high momenta. Unfortunately, the parameter range in which unambiguous results can be obtained is severely limited, due to theoretical and numerical constraints. 
The results of a pilot study using this method look compatible with the expected effect, but for the moment they do not allow us to draw definite conclusions about the large-$N$ limit; this aspect may be addressed again in future work.

In conclusion, we can say that the non-perturbative results obtained in the present work confirm the current theoretical understanding of this simple non-commutative model, and the virtues and limits of the fuzzy approach as a regularisation scheme. 

On one hand, it is clear that the effects associated with non-commutativity are intrinsic to the fuzzy regularisation, and the presence of the anomaly discussed above threatens the possibility to use fuzzy spaces as a straightforward regularisation for QFT in ordinary (\emph{i.e.} commutative) spaces. In this perspective, it would be particularly interesting to study in more detail the proposals that have been formulated~\cite{Dolan:2001gn, briandenjoe} to define an improved formulation of the action, yielding the correct QFT limit.

On the other hand, the results discussed in this paper provide evidence that fuzzy spaces are indeed a well-suited regularisation scheme for theoreies directly defined in non-commutative spaces --- e.g. Groenewold-Moyal $\R^n_\theta$ spaces --- and offer the possibility for a practical and efficient implementation of numerical studies of these models.

The success of this numerical study in the two-dimensional setting is very encouraging, and in the future it may be very interesting to address the $D=4$ case. This generalisation would obviously be of great interest from the physical point of view, and --- apart from a larger computational effort --- it is not expected to involve particularly difficult technical problems. In fact, the observation of the most interesting non-commutative effects may even turn out to be simpler than in $D=2$ --- due to the reasons discussed above.

\vskip1.0cm \noindent {\bf
Acknowledgements.}

\noindent It is a pleasure to thank A.P.~Balachandran, W.~Bietenholz, S.~Bourouaine, M.~Caselle, S.~Digal, B.P.~Dolan, F.~Garc\'{\i}a~Flores, T.R.~Govindarajan, K.S.~Gupta, M.~Hasenbusch, S.K.~Kar, X.~Martin, W.~Nahm, J.~Nishimura, D.~O'Connor, C.~S\"amann, H.~Steinacker and J.~Volkholz for enlightening discussions. The author's activity is currently supported through a Research Fellowship of the Alexander von Humboldt Foundation. A large part of the work presented here was done while the author was a postdoctoral research associate at the School of Theoretical Physics of the Dublin Institute for Advanced Studies (Ireland); the author acknowledges support received from Enterprise Ireland under the Basic Research Programme.

\newpage
\appendix{}
\vskip 0.5cm
\section{An overrelaxation algorithm for the fuzzy sphere}
\label{overrelsect}
\renewcommand{\theequation}{A.\arabic{equation}}
\setcounter{equation}{0}

In large parts of this work, 
a novel algorithm was used, that improves the efficiency of the numerical
simulation. It 
allows to damp the autocorrelation between subsequent elements in the Markov chain of configurations. 

The basic idea is related to the overrelaxation technique in lattice gauge theory~\cite{Adler:1981sn, Brown:1987rr}: 
The trial value in the update process of a given variable is chosen to be ``as far as possible'' from the original value, and such that the action of the system is unchanged.\footnote{This implies that the overrelaxation procedure is microcanonical; therefore the method is always combined with other canonical techniques, in order to ensure ergodicity of the whole update process.} In the lattice setting, this can be accomplished through a group reflection which can be worked out exactly for the $SU(2)$ group, and in an efficient way for a generic $SU(N)$ group~\cite{Petronzio:1990fb, deForcrand:2005xr}. This technique cannot be directly implemented in the present case, due to the fact that the $\Phi$ variable takes values in a domain of different nature: the space of hermitian matrices of size $N$ is non-compact, and, more important, a na\"{\i}ve ``reflection'' of the $\Phi$ matrix would not be effective for the purpose of reducing the autocorrelation time, since it would not allow to explore all of the physical orbits.
 
The algorithm we built generalises the principia underlying the overrelaxation technique, adapting them to the present case, and it
works as follows: Assume $\Phi_0$ to be the initial matrix configuration, obtained with some ergodic procedure; let $S_0=S(\Phi_0)$ be the associated euclidean action. Let $\Phi_\star$ be a new, completely random (and, therefore, completely independent from $\Phi_0$) hermitian matrix in Mat$_N$, with $S_\star=S(\Phi_\star)$ the corresponding value of the action. If $S_\star > S_0$,\footnote{If $S_\star \le S_0$, then $\Phi_\star$ is accepted as the new matrix configuration.} 
a new hermitian matrix $\Phi_1$, such that: $S_1=S(\Phi_1)=S_0$, can be built rescaling $\Phi_\star$ as:
\eq{rescaling}
\Phi_1 = \alpha \Phi_\star,
\en
provided the following condition: 
\eq{condition}
\left\{ (S_0>0)~\vee~\left( \tr \left( \Phi_\star \left[ L_i , \left[ L_i, \Phi_\star \right] \right] + r R^2 \Phi_\star^2 \right) < - R \sqrt{ -\frac{N \lambda S_0 }{\pi} \tr \Phi_\star^4 } \right) \right\}
\en
is true. If that is not the case, then $\Phi_\star$ is redefined (possibly iteratively) as:
\eq{newphistar}
\Phi_\star \longrightarrow \frac{\Phi_\star + \Phi_0}{2}
\en
until the condition in eq.~(\ref{condition}) is satisfied. Note that this shift would drive $\Phi_\star$ closer and closer to $\Phi_0$, thus inducing a correlation between corresponding matrix entries in $\Phi_1$ and  $\Phi_0$; nevertheless, in general the eventual value obtained for $\Phi_1$ may belong to a different physical orbit with respect to $\Phi_0$.

The algorithm is efficient under general conditions, including the cases in which $S(\Phi)$ is a function which varies strongly even for moderate changes in its argument, because the process driving $\Phi_\star$ towards $\Phi_0$ is exponentially fast, its implementation only involves trivial numerical operations, and terminates in a finite (and typically small) number of steps.\footnote{This is easily proven using continuity and the fact that the trivial $\alpha=1$ solution exists for $\Phi_\star = \Phi_0$.}

\begin{figure}
\centerline{\includegraphics[width=0.33\textwidth]{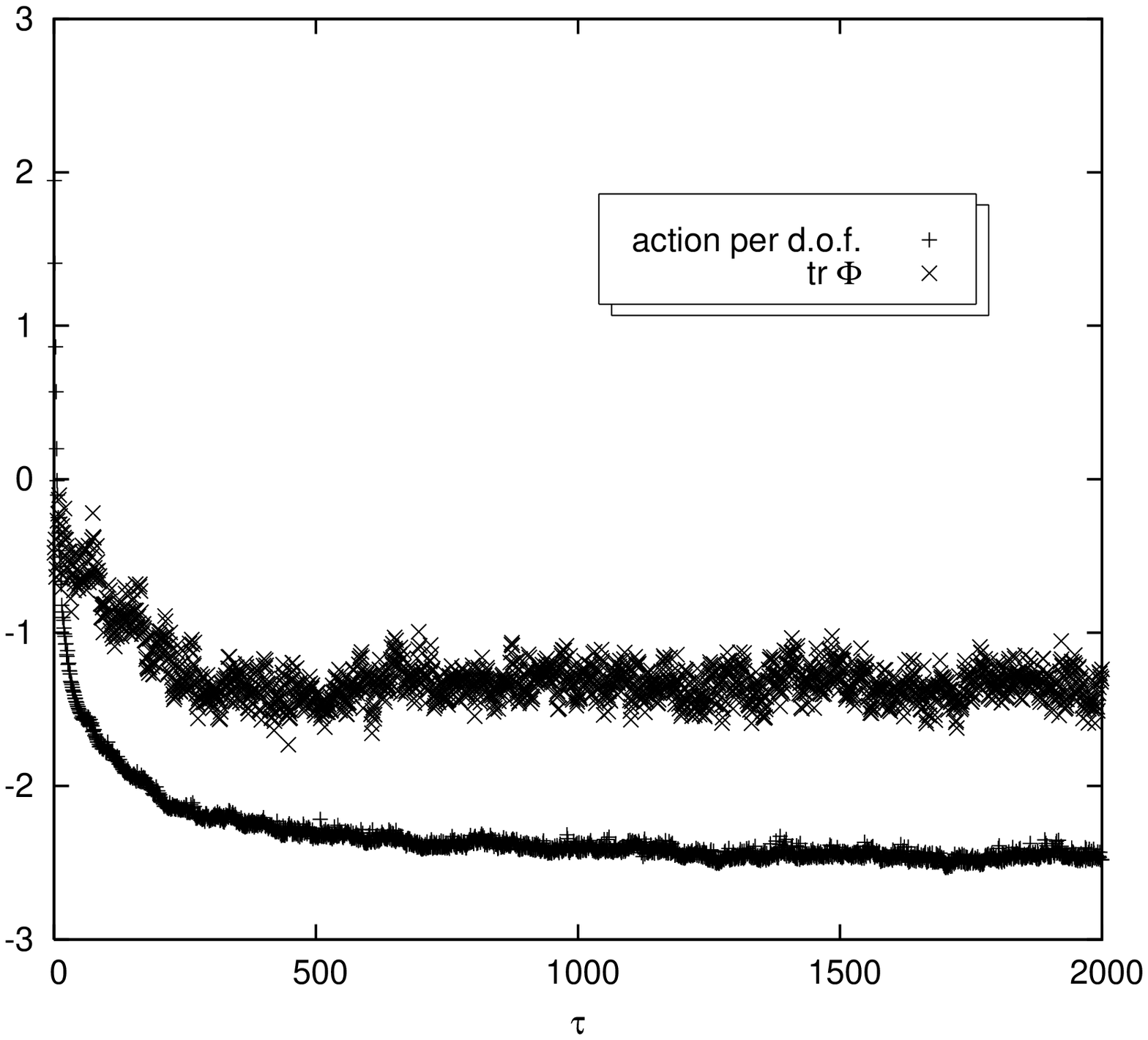}\includegraphics[width=0.33\textwidth]{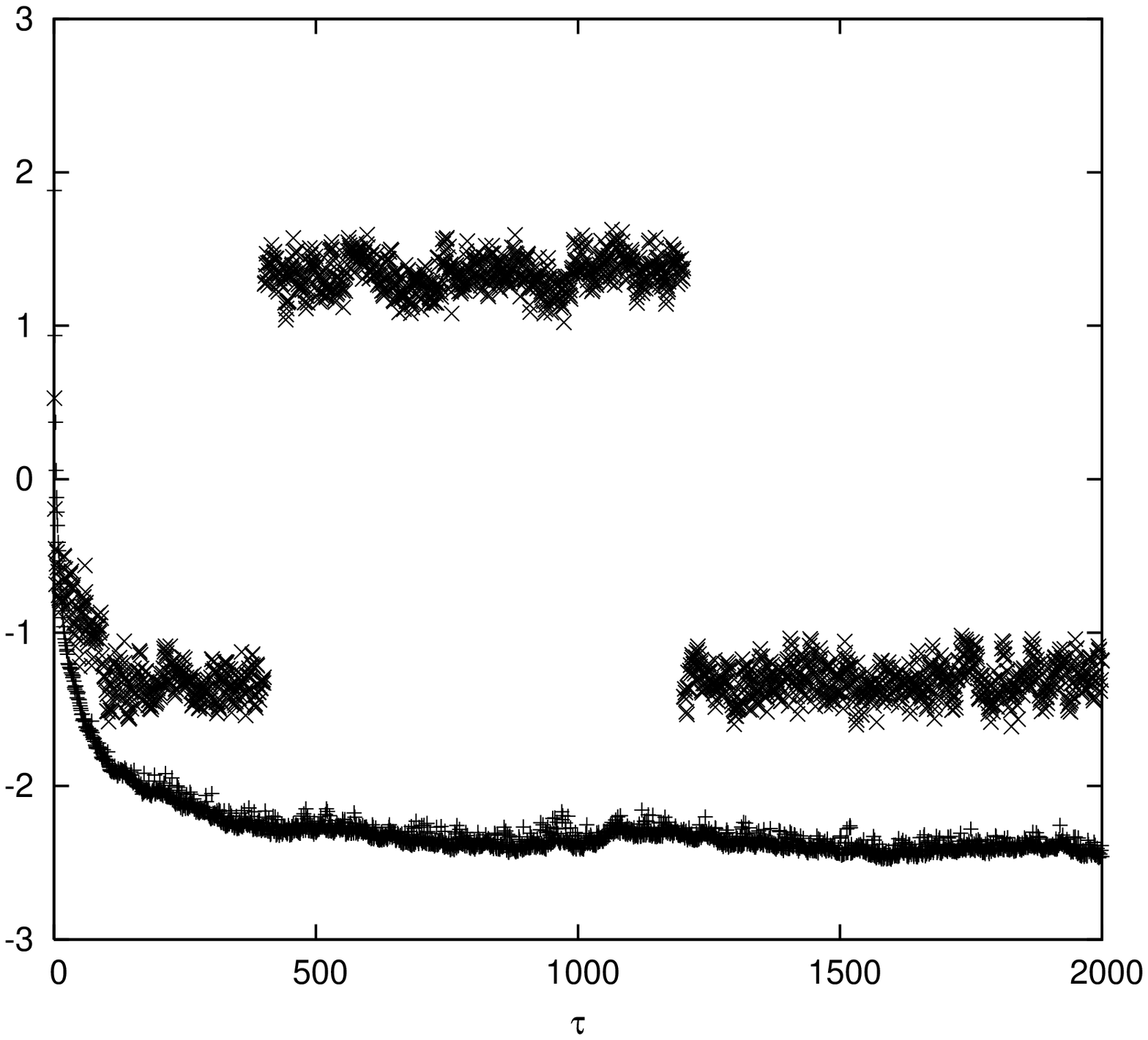}\includegraphics[width=0.33\textwidth]{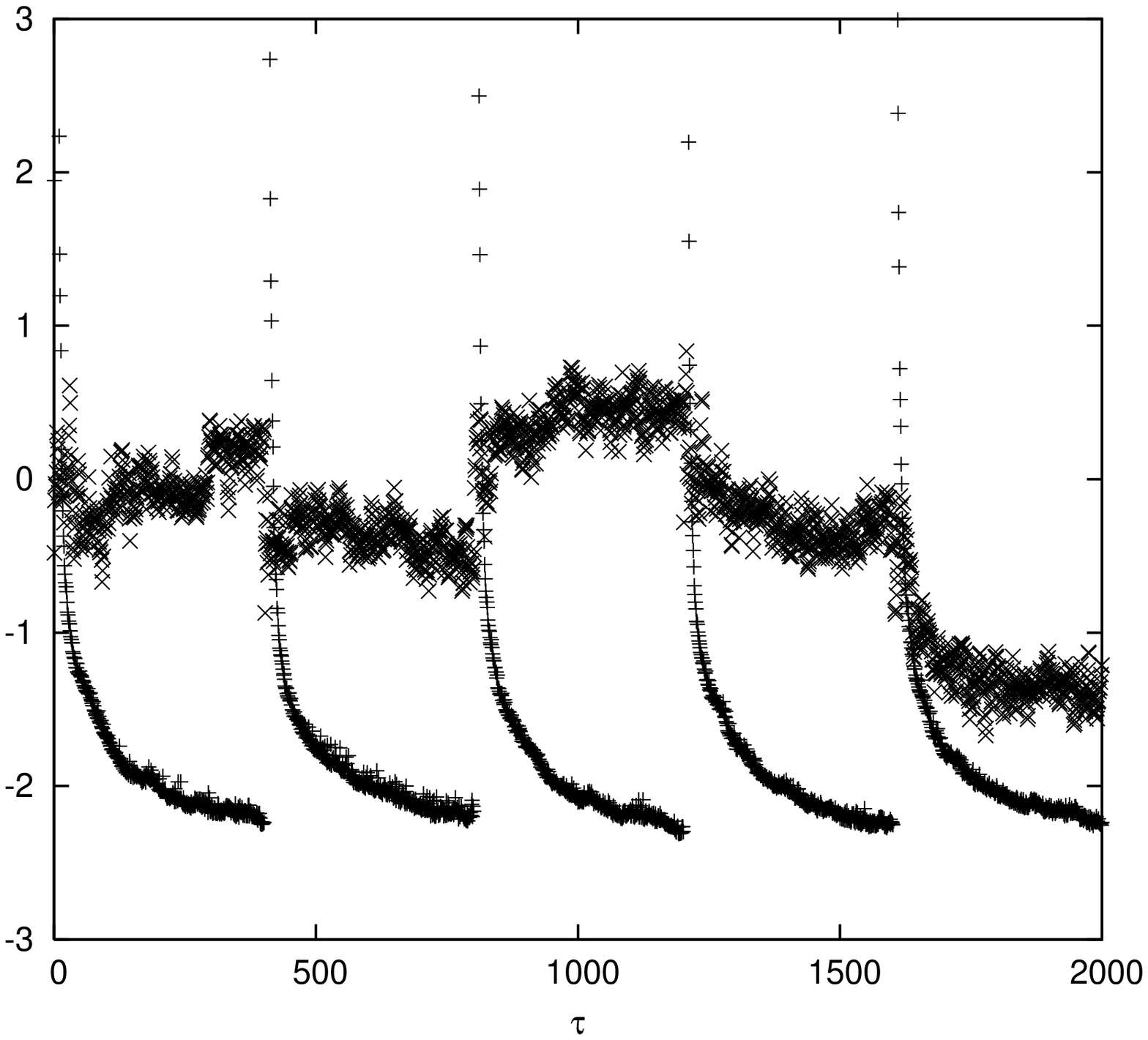}}
\caption{Left: A typical Monte Carlo history of the action per degree of freedom and of $\tr(\Phi)$, obtained without overrelaxation; the data fluctuate close to one of the two minima of the potential, and tunneling events are very rare. Data obtained from matrices of size $N=27$, in the two-cut regime. Center: Same as in the previous plot, but invoking the overrelaxation algorithm at Monte Carlo times $\tau = 400$, $800$, $1200$ and $1600$. Right: Same as in the previous plot, but replacing the overrelaxation steps with new starts from unthermalised configurations.}
\label{eff_fig}
\end{figure}

This algorithm proves superior to standard Metropolis (because it is not affected by the ergodicity problem) and to repeated new starts, because full thermalisation takes longer; figure~\ref{eff_fig} shows a comparison.

The initial hermitian matrix $\Phi_\star$ can be chosen according to an arbitrary distribution; in order to achieve the best efficiency, we tested various possibilities, and in the production runs the matrix elements of $\Phi_\star$ were chosen according to a gaussian distribution centered around zero; the width of the gaussian is tuned according to an optimisation criterion.

The number of ergodic updates between the overrelaxation steps is
another tunable parameter of the algorithm. Typically, for the results presented here, one overrelaxation step was invoked every 50, 200, 500, 1000, or 2000 steps.

\end{document}